\documentclass[11pt]{article}

\usepackage{acl}

\usepackage{times}
\usepackage{latexsym}

\usepackage[T1]{fontenc}
\usepackage[utf8]{inputenc}

\usepackage{microtype}

\usepackage{inconsolata}

\usepackage{graphicx}

\usepackage{tikz}
\usepackage{amsmath}
\usepackage{amssymb}
\usepackage{booktabs}

\usetikzlibrary{positioning}

\title{Predicting Oscar-Nominated Screenplays with Sentence Embeddings}

\author{Francis Gross \\
  Faculty of Informatics and Data Science \\
  University of Regensburg \\
  93053 Regensburg, Germany \\
  \texttt{francis.gross@stud.uni-regensburg.de}}

\begin{document}
\maketitle
\begin{abstract}
Oscar nominations are an important factor in the movie industry because they can boost both the visibility and the commercial success. This work explores whether it is possible to predict Oscar nominations for screenplays using modern language models. Since no suitable dataset was available, a new one called Movie-O-Label was created by combining the MovieSum collection of movie scripts with curated Oscar records. Each screenplay was represented by its title, Wikipedia summary, and full script. Long scripts were split into overlapping text chunks and encoded with the E5 sentence embedding model. Then, the screenplay embeddings were classified using a logistic regression model. The best results were achieved when three feature inputs related to screenplays (script, summary, and title) were combined. The best-performing model reached a macro-F1 score of 0.66, a precision recall AP of 0.445 with baseline 0.19 and a ROC-AUC of 0.79. The results suggest that even simple models based on modern text embeddings demonstrate good prediction performance and might be a starting point for future research.
\end{abstract}

\section{Introduction}

The "Oscar Bump", a well known phenomenon in the film industry, helps raising box-office\footnote{only cinema ticket-sales} numbers \citep{forbes2024}. Based on IBISWorld's data analysis \citep{bostonglobe2016} best picture nominees can receive about 22\% increase in box-office revenue after nomination and additionally about 15\% more after a win. The paper Oscarnomics \citep{oscarnomics} reports weekly grosses increase after nominations and wins are pronounced. Longer run-effects vary by title and content. Much more important today, Parrot Analytics \citep{parrot2025} shows that Best Picture nominations can also boost streaming revenues, underlining the growing relevance of digital distribution channels.

From the perspective of the film industry, screenplay nominations provide more than publicity and prestige. They also serve as a commercially relevant signal. Regardless of whether a script ultimately wins, the nomination process itself helps studios and producers benchmark projects and make better strategic investment decisions. For screenwriters, a nomination provides recognition and visibility that can significantly enhance their career opportunities.

Using AI to predict screenplay nominations could be beneficial for all involved parties, studios, producers, and writers alike. For studios and investors, such tools would help prioritize projects with higher nomination potential, reduce risk, and optimize investments early in development. For screenwriters, the promise of a nomination raises visibility, enhances negotiations, and may lead to more favorable opportunities.

In fact, Munich-based Leonine Studios \citep{leonine2024} has already integrated AI tools across creative development to improve productivity while keeping human writers central in the process. Similarly, companies like ScriptBook\citep{scriptbook2020} and Prescene \citep{prescene2024} offer AI-powered script analysis, feedback, and coverage that help evaluate narrative structure, pacing, and character consistency long before shooting begins.

These emerging practices suggest that there is substantial value in building predictive models for nominations, not just as research exercises but as practical systems with commercial, artistic and operational impact.

Previous research, such as Chiu et al. \citep{chiu2020screenplay}, has explored screenplay quality assessment using handcrafted features combined with standard classifiers. However, there has been little systematic work on leveraging modern long-text embeddings to directly predict award nominations from full scripts, summaries, and titles.

This paper presents a reproducible study of screenplay nomination prediction using a new dataset constructed from a movie script dataset combined with curated Oscar records. Main contributions of this work are: (1) construction of a merged dataset combining screenplay texts with Oscar nomination records, (2) evaluation of modern sentence embeddings, specifically the E5 embedding model  \citep{wang2022e5} with simple classifiers on representations of full scripts, summaries, and titles, and (3) discussion of potential implications of such models for research and industry practice.

\section{Related Work}

Film revenues are well known to be often influenced by the amount of media attention a movie can generate. Because of this, researchers have studied different ways to predict award outcomes to understand their role in investment decisions. Some early works use economic and statistical models to predict Academy Award winners from historical data and other features \citep{pardoe2005,pardoe2007}. 

Other studies include social signals, for example, data from online forums or social networks, to see if audience attention can help predict both box office performance and award outcomes \citep{krauss2008}. 

There is also work that looks directly at the scripts. Chiu et al. \citep{chiu2020screenplay} proposed a model that combines handcrafted features, such as dialogue ratio or sentiment, with standard classifiers, as already mentioned. They report that this can predict screenplay nominations to some degree, but it still needs a lot of feature engineering. 

More recent work on the MovieSum dataset \citep{chitale2025_moviesum} shows that modern long-text embeddings can be applied to full screenplays and summaries, but these studies focus on summarization, e.g. for automatic coverage writing. 

This work builds on these ideas. In contrast to earlier approaches, sentence embeddings are used as the central representation, and simple classifiers are tested on them. In this way it can be studied whether embeddings alone are able to provide strong performance on the task of predicting Oscar nominations for screenplays.
% ---------------- Methods starting ----------------------------------------------
\section{Methods}

\paragraph{Solving the Issue.} 

\begin{figure}[htbp]
\centering
\begin{tikzpicture}[
  node distance=8mm,
  every node/.style={font=\small, align=center},
  box/.style={draw, rounded corners, minimum width=4.5cm, minimum height=1cm, fill=gray!10}
]

% Inputs combined
\node[box] (inputs) {Screenplay inputs: \\ Title (1), Summary ($\sim$2--6), Script ($\sim$20--186) chunks};

% Pooling
\node[box, below=of inputs] (pooling) {E5 encoding + Mean/Max pooling \\ $\rightarrow$ per-field vectors};

% Fusion
\node[box, below=of pooling] (fusion) {Concatenation of vectors \\ (Title, Summary, Script)};

% Classifier
\node[box, below=of fusion] (cls) {Logistic Regression \\ (nomination prediction)};

% Arrows
\draw[->] (inputs) -- (pooling);
\draw[->] (pooling) -- (fusion);
\draw[->] (fusion) -- (cls);

\end{tikzpicture}
\caption{Compact task pipeline.}
\label{fig:embedding_pipeline_compact}
\end{figure}
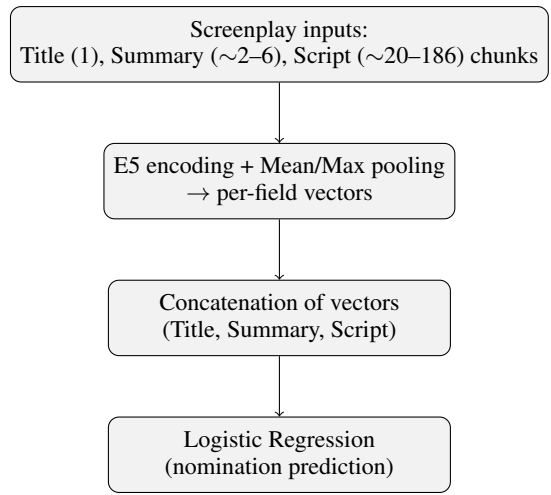

Since there was no appropriate dataset available, a new one was created from other existing data. Another issue was that movie scripts have on average more than 30k tokens, they cannot be used in transformer encoders directly. To solve this problem, each screenplay was prepared and split into overlapping chunks, which were independently encoded with E5. For classification a logistic regression model was applied. A compact task pipeline overview is shown in Figure~\ref{fig:embedding_pipeline_compact}.

% grafik pipeline

% grafik ende

\subsection{Data}

Screenplays are the foundation of the model of this work, since they must be officially submitted to the Academy in order to receive an Oscar nomination \citep{academy_rules}. This makes the screenplay the central source of information in the nomination process. Assuming that voters have read the screenplays, at least the summaries, but almost certainly noticed the title, following features were taken into account: \textbf{script text}, \textbf{summary} and \textbf{title}.

These features were extracted from MovieSum \citep{saxena2024moviesum}, a labeled dataset of 2,200 movie screenplays with corresponding Wikipedia summaries. The original screenplay texts are manually formatted with xml-tags representing their structural elements. Additionally the movie names are matched with their IMDb ID\footnote{Internet Movie Database (IMDb) unique identifier} for collecting metadata if necessary or merging with other datasets. MovieSum is available as a Hugging Face dataset.

The labels, whether or not a movie received a nomination, were taken from the curated Oscar records from GitHub \citep{oscardata2022}, which contain categorical Academy Award nominations and winners, starting from 1927 to 2024, linked to their IMDb IDs. A movie that could be found in this records definitely was nominated. The records also show if a nominated film has won. For selecting appropriate categories of screenplays awarded, this work used the records classes "Writing" and "Title". These are meta-labels in the Oscar records corresponding to the tasks of this work. 

A new dataset, called \textbf{Movie-O-Label} \citep{gross2025movieolabel}, was constructed by merging MovieSum with the Oscar records using the IMDb ID as key. It contains 2,200 entries, of which around 19\% are positive (nominated). This imbalance reflects the real-world rarity of Oscar nominations, where only 
few screenplays are awarded each year in the category "Writing".

From the MovieSum dataset the fields \texttt{movie\_name}, \texttt{imdb\_id}, \texttt{script}, and \texttt{summary} were taken as is. The \texttt{movie\_name} follows the format name\_YYYY, from which the \texttt{title} (film name without year) and the \texttt{year} were split. The \texttt{script} field contains the full screenplay structured with xml-tags. From this derive two additional script text versions: \texttt{script\_plain}, where only xml-tags are removed, and \texttt{script\_clean}, where regex-based cleaning rules were applied to \texttt{script\_plain} to remove artifacts such as transitions (e.g. scene transition like \textit{CUT TO}), and normalize unicode characters. This results in cleaner text while keeping the original script structure and semantics. The \texttt{summary} field contains the Wikipedia plot synopsis for each film as already mentioned.

From the Oscar records derive the fields \texttt{nominated} and \texttt{winner}, which were assigned by matching via \texttt{imdb\_id}. A film with a nomination in the "Writing" or "Title" class is labeled \texttt{nominated=1}, all others as \texttt{nominated=0}. The field \texttt{winner} is set to 1 only for winning screenplays. Combining these sources constructs the final dataset Movie-O-Label. An overview of the variables stored is shown in Table~\ref{tab:movieolabel_columns}.

\begin{table}[h]
\centering
\small
\begin{tabular}{p{0.28\linewidth} p{0.62\linewidth}}
\hline
\textbf{Column} & \textbf{Description} \\
\hline
\texttt{movie\_name} & Original identifier \textit{title\_YYYY} \\
\texttt{imdb\_id} & IMDb ID for merging datasets \\
\texttt{script} & Raw script with xml-tags \\
\texttt{script\_plain} & script without xml-tags \\
\texttt{script\_clean} & Normalized and cleaned script \\
\texttt{summary} & Wikipedia movie summary \\
\texttt{title} & Film title from \texttt{movie\_name}  \\
\texttt{year} & Release year from \texttt{movie\_name} \\
\texttt{nominated} & Label (1 = nominated, 0 = not nominated) \\
\texttt{winner} & Label (1 = win, 0 = otherwise) \\
\hline
\end{tabular}
\caption{Columns in the new Movie-O-Label dataset.}
\label{tab:movieolabel_columns}
\end{table}

For training and evaluation, the dataset was split into training, validation, and test 
sets with a ratio of 60/20/20. An overview of the label distribution is shown in Table~\ref{tab:dataset_stats}. The ratio was chosen to provide enough validation data for threshold tuning while not reducing the training set too much. The splits are stratified by the label to preserve the class distribution and stored in a numpy zip file ( \texttt{.npz}).

 \begin{table}[t]
\centering
\small
\begin{tabular}{lccc}
\hline
\textbf{Split} & \textbf{Total} & \textbf{Label=1} & \textbf{Label=0} \\
\hline
Train (60\%)       & 1320 & 250 \,(18.94\%) & 1070 \,(81.06\%) \\
Val (20\%)  &  440 &  84 \,(19.09\%) &  356 \,(80.91\%) \\
Test (20\%)        &  440 &  83 \,(18.86\%) &  357 \,(81.14\%) \\
\hline
\textbf{Total}     & 2200 & 417 \,(18.95\%) & 1783 \,(81.05\%) \\
\hline
\end{tabular}
\caption{Label distribution in the Movie-O-Label dataset (nominated=1, not nominated=0).}
\label{tab:dataset_stats}
\end{table}

 Token statistics for the three features using the E5 tokenizer are shown in Table~\ref{tab:token_stats}. Titles are very short, with a median of 3 tokens. Summaries have a median of about 826 tokens and are already relatively long. Scripts are very long with a median length of over 31k tokens and a maximum of 97k tokens. These numbers show the challenge of modeling full screenplays directly, as their length exceeds the input limits of transformer models and would result in very sparse, high-dimensional representations for TF-IDF models.

\begin{table}[h]
\centering
\small
\begin{tabular}{lrrrrr}
\hline
\textbf{Field} & \textbf{Median} & \textbf{Mean} & \textbf{Min} & \textbf{Max} \\
\hline
Title   & 3     & 3.2    & 1    & 17 \\
Summary & 826   & 797.3   & 14   & 2311 \\
Script  & 31898 & 32455.5 & 5891 & 97086 \\
\hline
\end{tabular}
\caption{Token lengths (E5 tokenizer) for titles, summaries, and scripts 
in the Movie-O-Label dataset.}
\label{tab:token_stats}
\end{table}

\subsection{Solution Approach and Methodology}

\paragraph{Chunking.}

The selected features for the model, taken from Movie-O-Label, were \texttt{script\_clean}, \texttt{summary} and \texttt{title}. The feature texts of \texttt{script\_clean} and \texttt{summary}, were split into overlapping word-based chunks of max. 400 words with an overlap of 80 words to preserve context. The short \texttt{title} texts were processed as single chunks. The chunking statistics are shown in Table~\ref{tab:chunkstats}.

\begin{table}[h]
\centering
\small
\begin{tabular}{lccc}
\hline
Field & Avg.\ Chunks & Max & Docs \\
\hline
Title   & 1.0  & 1   & 2200 \\
Summary & 2.5  & 6   & 2200 \\
Script  & 77.0 & 186 & 2200 \\
\hline
\end{tabular}
\caption{Chunk statistics.}
\label{tab:chunkstats}
\end{table}

\paragraph{Embeddings.}
All chunk embeddings were encoded with E5, the pre-trained e5-base-v2 model \citep{e5modelcard}, using its SentenceTransformer. As recommended for classification, the input prefix \texttt{"query: "} was applied. E5 was chosen because it is lightweight enough for the hardware and GPU used, publicly available, has an easy prompting style and provides strong semantic representations for the binary classification task of this work. Respective to the available GPU the embeddings were computed in mini-batches with batch\_size = 96 for script and summary chunks, while a larger batch\_size = 256, was possible for the short title texts.

Per screenplay each text chunk was encoded into an embedding vector $\mathbf{h}_i \in \mathbb{R}^d$. Where $d$ is the embedding dimension, with $d{=}768$ for \texttt{e5-base-v2} and $i$ is the index of a chunk. Where $n$ is the number of chunks for a document, Mean-Max-Pooling was computed.

Mean pooling calculates the averages of all embeddings of a document. Intuitively, this creates a single vector that represents the overall meaning of a script. Every chunk contributes equally, so no single scene dominates. The mean embeddings vector represents a global summary of the script’s semantic.
\begin{align*}
\mathbf{mean} &= \frac{1}{n}\sum_{i=1}^n \mathbf{h}_i
    &&\text{(mean pooling)}\\[4pt]
\end{align*}

Max pooling takes, for each embedding dimension, the maximum value across all chunk embeddings. Intuitively, this keeps the strongest semantic signals that appear anywhere in the script. So, the important features are preserved in the final vector even if they would be diluted by mean-pooling averaging.
\begin{align*}
\mathbf{MVec} &= \max_{1\le i\le n}\mathbf{h}_i
    &&\text{(max pooling)}\\[4pt]
\end{align*}

After Mean-Max-Pooling a script is represented by a concatinated vector $v$ as follows:
\begin{align*}
\mathbf{v} &= [\mathbf{m};\mathbf{MVec}]
    &&\in\mathbb{R}^{2d}\\[4pt]
\end{align*}
Finally a standardized vector $\hat{v}$ is computed by the Euclidian L2 normalization. This step removes differences in vector length and keeps only the direction of the embedding. This makes the features comparable in scale and stabilizes the logistic regression classifiers.
\begin{align*}
\hat{\mathbf{v}} &= \frac{\mathbf{v}}{\lVert\mathbf{v}\rVert_2}
    &&\text{(L2-normalization)}
\end{align*} \\

After pooling and normalization, the field vectors (title, summary, script) were concatenated using a vertical stack operation (\texttt{numpy.vstack}) to produce the final screenplay embedding for the models with combined features. The resulting matrix was stored with \texttt{joblib} for reuse during training and evaluation.

\paragraph{Classification.}
After embedding, the cached matrices are used to fit a logistic regression classifier. This model was assumed to be fully sufficient for the task of this work, because the main semantic and structural information is already captured by the pre-trained text embeddings. The classifier only should need to learn a simple linear decision boundary.

\begin{table}[h]
\centering
\small
\begin{tabular}{lp{4cm}}
\hline
\textbf{Parameter} & \textbf{Value} \\
\hline
Classifier & Logistic Regression \\
Penalty & L2 \\
Solver & lbfgs \\
Reg. $C$ & 1.0 \\
Max iters & 5000 \\
Class weight & balanced \\
Chunk size & 400 w, 80 ovlp \\
Embeddings & E5-base-v2 \\
Pooling & Mean + Max, L2-norm \\
Feature fusion & Script / Summary / Title \\
Split & 60/20/20 strat. \\
Thresh.\ opt. & $\tau^\ast$: max F1$_\mathrm{pos}$ (Val) \\
\hline
\end{tabular}
\caption{Key hyperparameters for logistic regression.}
\label{tab:hyperparams}
\end{table}

\begin{table*}[t]
\centering
%\small
\begin{tabular}{lcccccc}
\hline
Variant & Acc & ROC & PR & F1$_\text{pos}$ & F1$_\text{neg}$ & MacroF1 \\
\hline
Script+Summary+Title & \textbf{0.759} & \textbf{0.790} & \textbf{0.455} & \textbf{0.485} & \textbf{0.843} & \textbf{0.664} \\
Script+Summary       & 0.734 & 0.768 & 0.450 & 0.485 & 0.821 & 0.653 \\
Summary              & 0.693 & 0.754 & 0.424 & 0.444 & 0.788 & 0.616 \\
Script               & 0.732 & 0.742 & 0.415 & 0.427 & 0.825 & 0.626 \\
Title                & 0.670 & 0.692 & 0.347 & 0.388 & 0.774 & 0.581 \\
\hline
\end{tabular}
\caption{Test performance of the logistic regression models.  
ROC = ROC-AUC, PR = PR-AUC.}
\label{tab:results}
\end{table*}

The main training and model settings are summarized in Table~\ref{tab:hyperparams}.
Most of the hyperparameter were set to the default configuration of scikit-learn. This fits already the task of this work well. L2 regularization is the standard and helps keep weights small when using high-dimensional embeddings. The lbfgs solver is efficient for L2 problems. First, the parameter \texttt{max\_iter} was set to 5000 because the default was not sufficient for convergence with our long feature embeddings. This should guarantee that the optimizer can reach a stable loss minimum. Second, the parameter \texttt{class\_weight} was set to "balanced" because our dataset is imbalanced (only about 19 \% nominated). This should improve recall and F1 score. The decision threshold $\tau^\ast$ was chosen on the validation set by scanning thresholds from 0.05 to 0.95 and selecting the value that maximized the positive-class F1 score. 

\paragraph{Model Variants.}
The logistic regression classifier was trained and tested on five feature configurations:
\begin{itemize}
\item \textbf{TITLE}: \texttt{title} embedding only.
\item \textbf{SCRIPT}: \texttt{script\_clean} embedding only.
\item \textbf{SUMMARY}: Wikipedia \texttt{summary} embedding only.
\item \textbf{SCRIPT+SUMMARY}: concatenation of \texttt{script\_clean} and \texttt{summary} embeddings.
\item \textbf{SCRIPT+SUMMARY+TITLE}: concatenation of \texttt{script\_clean}, \texttt{summary}, and \texttt{title} embeddings.
\end{itemize}

For all configurations the fixed train-test split file was used, also the same hyperparameter.

\subsection{Implementation and Reproducibility}
The model is implemented in Python using Jupyter Lab and common data science, open-source libraries. For calculating sentence embeddings, PyTorch GPU and SentenceTransfomers were applied. Everything ran on a PC with an NVIDIA GeForce RTX 3060 GPU (12 GB VRAM). An embeddings calculation run took about 1 hour. 

For reproducibility the python code, fixed data splits, saved embeddings and a JSON file with threshold and model parameters, are provided. Everything is available on Hugging Face as additional dataset files \citep{gross2025movieolabel}.

\section{Results}

Since the dataset is imbalanced, only about 19\% with label nominated, this work reports the macro F1 score and the PR-AUC in addition to Accuracy and ROC-AUC. Macro F1 treats the positive and negative classes equally, so it indicates better when the positives are rare. The PR-AUC is useful on imbalanced datasets because it focuses on the positive class and ignores true negatives. Although the model of this work and the dataset are not directly comparable to the work of \citet{chiu2020screenplay}, because they used different datasets with manually engineered features based on narrative structure, it is interesting that simple embedding-based approaches reach a similar performance level in terms of ROC-AUC (around 0.79 vs.\ $\sim$0.75 reported by Chiu et al.).

The model performance results of this work are shown in Table~\ref{tab:results} for all variants of features. The best performance was achieved by the \textbf{ script + summary + title} model with Accuracy~0.76, ROC-AUC~0.79, PR-AUC~0.455, Macro-F1~0.66.  

Adding both the full script and the Wikipedia summary clearly improves over using only a single feature. Titles alone are weak predictors (ROC-AUC~0.69, Macro-F1~0.58), but still better than random.
Combining script and summary already show much better results (ROC-AUC~0.77, PR-AUC~0.45). 

\begin{figure}[tbh]
    \centering
    \includegraphics[width=\columnwidth]{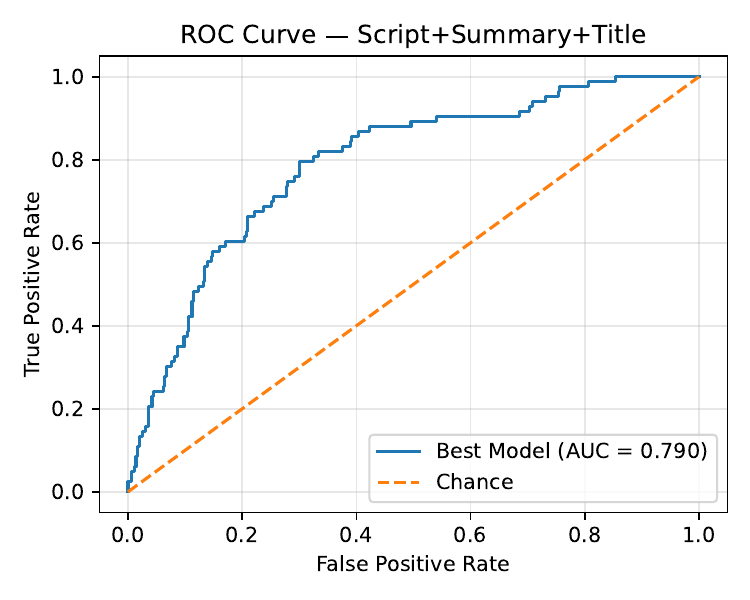}
    \caption{ROC-AUC of the best-performing model, Script+Summary+Title.}
    \label{fig:roc}
\end{figure}

\begin{figure}[tbh]
  \centering
  \includegraphics[width=\linewidth]{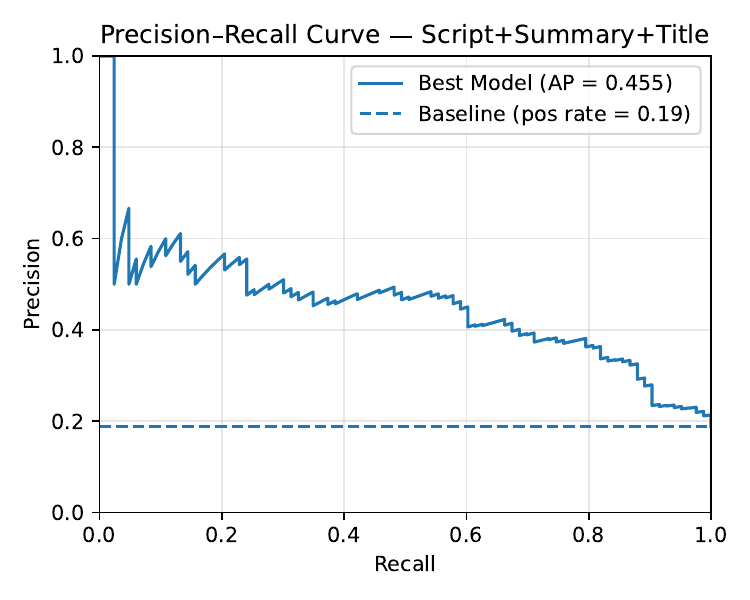}
  \caption{PR-AUC of the best-performing model, Script+Summary+Title.}
  \label{fig:pr}
\end{figure}

ROC and PR curve (Fig.~\ref{fig:roc},~\ref{fig:pr}) show that the best-performing model dominates across thresholds. The curves stay clearly above the random baselines, and the areas under the curves confirm strong separation. A random classifier would reach a PR-AUC equal to the positive class rate (about 0.19 in the dataset). The best-performing model achieves a PR-AUC of 0.455, which shows that it can identify nominated scripts much better than chance, even with the strong class imbalance.

\section{Discussion}

The results show that sentence embeddings combined with a linear classifier already can differentiate between nominated and non-nominated screenplays. Most of the predictive power probably comes from the text embeddings and not from the classifier itself. So a simple logistic regression is fully sufficient. Using both script and summary outperforms single feature models. Adding the title gives a small improvement of model performance. Because the dataset is highly imbalanced, PR-AUC and macro-F1 are the most informative metrics in addition to ROC-AUC. PR-AUC summarizes precision–recall trade-offs for the positive class and is recommended on imbalanced data, whereas ROC-AUC can be too optimistic applied on imbalanced datasets \citep{saito2015}.

Adding more textual information clearly helps. Using only the movie title performs poorly, while using the full script or the Wikipedia summary improves the model much more. Combining script and summary achieves stronger results. This suggests that plot descriptions and detailed scene-level text contain complementary signals about Oscar potential as was expected in this work. 

Despite reasonable predictive scores, the model is far from perfect. Certainly Oscar nominations for screenplays depend on many non-textual factors, too. For example such as production campaigns, studio reputation, or release timing, which  were not considered here. Since Oscar nominations are decided by human voters their decisions can be influenced by many subjective factors, e.g. personal interests, current industry trends, social dynamics. Also, voters can have personal preferences, dislikes, or loyalties toward certain writers, studios, or creative teams. These aspects are not represented in the screenplay text and therefore remain outside the scope of this model. Likewise, real human voters have to review the movies by watching them, it is absolutely recommended for building better models to consider video to LLM. 

The dataset itself is limited in size and coverage, because not all Oscar-nominated screenplays are available in the MovieSum collection. In addition, the Wikipedia summaries included in the dataset vary in length and quality, which may add noise to the features. Another limitation is the focus on Oscar nominations only. For practical investment decisions of production companies, it might be more useful to predict multiple major screenplay awards, not just the Oscars, e.g. Golden Globes, BAFTA, César, Lola etc. A further limitation is that this work only considers English-language screenplays. But for the international film business it could be more interesting to develop more multilingual models.

Hardware limitations constrained this work to use larger encoders. Pre-tests with e5-large-v2 embeddings  could not outperform the best-performance model of this work. It is suspected that the used dataset (only 2200 scripts) is too small for the larger embedding space and that chunking long scripts into 400-word segments may have limited the advantage of larger pretrained model in addition to the hardware constraints, e.g. half-precision floating point FP16 and just 12GB VRAM.

\section{Conclusion and Future Work}

This work shows that it is possible to build a simple pipeline to predict whether a screenplay will be nominated for an Academy Award - Oscars. For other film awards the same approach could be easily adapted, since it only requires screenplay texts and reliable nomination records.  

The available hardware of this approach is modest and the code can run on a single mid-range GPU (e.g. RTX 3060, 12 GB VRAM). This means that the method is not expensive or technically demanding and could even be used by individual screenwriters or small production teams. So, they can receive an early signal about the potential of their scripts and this may help for improvement.

For the film industry, this work contributes just a small component in the broader field of screenplay analysis and decision support. Since current state of the art algorithms used in commercial tools are proprietary and undisclosed, this work cannot directly compare to existing industry solutions. But for the film industry script analysis and evaluation tools exist and are suggested to be used in development and production workflows \citep{forbes2024_ai_cinema}. 

\textbf{Future work} could explore several directions. First, using larger language models with high performance hardware. This also could reduce hard chunking. Second, integrating \textbf{video-to-LLM} signals, e.g. from trailers or released films. This could improve the model's predictive ability by providing complementary information that pure text can not capture.Third, the dataset and the model should be extended to other awards than the Oscars. This could be interesting for both, screenwriters and industry. Forth, combining other non-text features such as budget, genre, studio, or marketing signals with the embedding model might capture important nomination factors beyond the script itself. Finally, an artificial jury could be created from LLM personas. So, the human voting process can probably be simulated and improve model predictions. This could be built on prior work on LLM-as-a-Personalized-Judge which shows potential but also warns about reliability issues \citep{llm_personalized_judge_2024}. While these works are not specific to screenplay evaluation, they can provide a starting point for building and validation LLM persona-based movie awarding juries.

\bibliography{custom}

\end{document}